\documentclass[sigconf]{acmart}
\usepackage{pifont}
\renewcommand\footnotetextcopyrightpermission[1]{} 
\AtBeginDocument{%
  \providecommand\BibTeX{{%
    \normalfont B\kern-0.5em{\scshape i\kern-0.25em b}\kern-0.8em\TeX}}}

\setcopyright{acmcopyright}
\copyrightyear{2018}
\acmYear{2018}
\acmDOI{10.1145/1122445.1122456}

\acmConference[Woodstock '18]{Woodstock '18: ACM Symposium on Neural
  Gaze Detection}{June 03--05, 2018}{Woodstock, NY}
\acmBooktitle{Woodstock '18: ACM Symposium on Neural Gaze Detection,
  June 03--05, 2018, Woodstock, NY}
\acmPrice{15.00}
\acmISBN{978-1-4503-9999-9/18/06}

\begin{document}

\title{Characterizing Differentially-Private Techniques in the Era of Internet-of-Vehicles}

\author{Yicun Duan*, Junyu Liu*,  Wangkai Jin, Xiangjun Peng}
\affiliation{
\institution{User-Centric Computing Group, University of Nottingham Ningbo China}
\url{https://unnc-ucc.github.io}
}


\thanks{* stands for equal contribution.}
\renewcommand{\shortauthors}{Wang, et al.}

\newcommand{\concept}{\textit{Internet-of-Vehicles}}
\newcommand{\conceptabbr}{\textit{IoV}}

\newcommand{\DP}{\textit{Differential Privacy}}
\newcommand{\DPabbr}{\textit{DP}}

\begin{abstract}
Recent developments of advanced Human-Vehicle Interactions rely on the concept \concept{} (\conceptabbr{}), to achieve large-scale communications and synchronizations of data in practice. The concept of \conceptabbr{} is highly similar to a distributed system, where each vehicle is considered as a node and all nodes are grouped with a centralized server. In this manner, the concerns of data privacy are significant since all vehicles collect, process and share personal statistics (e.g. multi-modal, driving statuses and etc.). Therefore, it's important to understand how modern privacy-preserving techniques suit for \conceptabbr{}.

We present the most comprehensive study to characterize modern privacy-preserving techniques for \conceptabbr{} to date. We focus on \DP{} (\DPabbr{}), a representative set of mathematically-guaranteed mechanisms for both privacy-preserving processing and sharing on sensitive data. The purpose of our study is to demystify the tradeoffs of deploying \DPabbr{} techniques, in terms of service quality. We first characterize representative privacy-preserving processing mechanisms, enabled by advanced \DPabbr{} approaches. Then we perform a detailed study of an emerging in-vehicle, Deep-Neural-Network-driven application, and study the upsides and downsides of \DPabbr{} for diverse types of data streams. Our study obtains 11 key findings and we highlight FIVE most significant observations from our detailed characterizations. We conclude that there are a large volume of challenges and opportunities for future studies, by enabling privacy-preserving \conceptabbr{} with low overheads for service quality.

\end{abstract}


\maketitle

\pagestyle{headings}
\setcounter{page}{1}
\pagenumbering{arabic}
 
\section{Introduction}
Techniques for Human-Vehicle Interactions naturally evolve from \textit{Vehicle Ad-hoc Networks (VANETs)} to \concept{} (\conceptabbr{}), to realize frequent, large-scale and efficient communications and synchronizations of data in practice. Unlike \textit{VANETs}, \conceptabbr{} views multiple vehicles as nodes within a distributed system, where groups of nodes are powered by a centralized server for rich computational resources to achieve diverse types of usages. For instance, transportation scheduling requires detailed vehicles information gathering (e.g. coordinates, speed and etc.), and performs the scheduling according to the collected statistics \cite{traffic-scheduling,traffic_control}. Similar usages, via \conceptabbr{}, become more imminent and important with the emerging trends of Autonomous Vehicles. However, though \conceptabbr{} provides extensive opportunities for various novel techniques in Human-Vehicle Interaction (e.g. healthcare applications~\cite{healthcare-app}, lane-changing warnings~\cite{lane-changing,lane-changing2} and etc.~\cite{SHARMA2019100182}), it also significantly raises the concerns of privacy in practice. Hereby, We summarize them into the following two aspects.

First, sophisticated designs for better user experiences are usually composed with high demands for computational resources (i.e. centralized servers in \conceptabbr{}), and this results in a large amount of sensitive data being exposed outside the vehicles. For instance, recent advances of Deep Neural Networks provide rich functionalities (e.g. object detection \cite{object-det}, object recognition \cite{object-recog} and etc.) but demand a huge amount of computational resources \cite{sze2017efficient}. In \conceptabbr{}, sensitive data is possible to be gathered in centralized servers for processing purposes, which incurs the privacy leakage in potential.

Second, the design pre-requisites of several key applications, in the context of \conceptabbr{}, demand all vehicles to share data within centralized servers. Such action enforces all vehicles to submit sensitive data for centralized data sharing and processing. For instance, recent developments take advantage of dash recordings to identify the root causes of vehicle accidents, but require all dash recordings to be gathered in centralized servers \cite{dashcam}. In this manner, gathered sensitive data are highly likely to become the target of adversarial attacks, which would impair the end-users‘ privacy integrity when disclosed to the untrusted parties.

Therefore, it's important to introduce privacy-preserving techniques in the era of \conceptabbr{}. However, since no prior works systematically to identify and tackle this issue, it's still unclear how modern privacy-preserving techniques perform in the context of \conceptabbr{}. In this work, we focus on \DP{} (\DPabbr{}), a representative set of mathematically-guaranteed mechanisms, because \DPabbr{} can enable both privacy-preserving processing and sharing on sensitive data. 

Our goal in this work is to investigate representative mechanisms of \DPabbr{} for both processing and sharing on sensitive data, and experimentally characterize the advantages and disadvantages in the context of \conceptabbr{}. To this end, we design a rigorous methodology to understand the impacts of \DPabbr{}, for processing and sharing respectively. For privacy-preserving processing in \conceptabbr{}, we abstract the interactions between centralized servers and vehicle nodes and characterize the effects from \DPabbr{}-protected processing of sensitive data; and for privacy-preserving sharing in \conceptabbr{}, we select a representative in-vehicle application and characterize the impacts of \DPabbr{}-enabled sharing of sensitive data. (Section~\ref{sec:methodology})

We first focus on \DPabbr{}-enabled processing for sensitive data, by abstracting the interactions between vehicle nodes and centralized servers. For \DPabbr{}-enabled processing, centralized servers perform queries to retrieve sensitive data from vehicle nodes, and then perform the computation within servers. In our characterizations, we select FOUR representative mechanisms for \DPabbr{}-enabled processing, and design representative workloads according to these mechanisms. Then we characterize the effects of these techniques and analyze the results comprehensively, in terms of absolute/relative errors and service quality (Section~\ref{sec:processing}).

We then focus on \DPabbr{}-enabled sharing for sensitive data, by leveraging a representative in-vehicle application (i.e., a state-of-the-art in-vehicle predictor for drivers' statistics taking facial expressions as the only input) using Deep Neural Networks (DNN). In our characterizations, we consider both centralized and distributed training of these DNN models, by enabling \DPabbr{} on diverse types of input data streams and the trade-off between two DP protection mechanisms (i.e. Global \DP{} and Local \DP{}) with different privacy-preserving scales. Then we characterize the effects of different parameter setting combinations of these techniques and analyze the results comprehensively, with a particular focus on service quality (Section~\ref{sec:sharing}).

Derived from the above two large-scale studies, we make 11 key observations (i.e. distributed into separate sections) and highlight FIVE most significant observations here: \ding{202} \textbf{there are no one-size-fits-all mechanisms of \DPabbr{} for all applications of \conceptabbr{}, in terms of Relative Error}; \ding{203} \textbf{there are no one-size-fits-all mechanisms of \DPabbr{} for all applications of \conceptabbr{}, in terms of Absolute Error as well}; \ding{204} \textbf{to strike a balance between data integrity and privacy preserving, a preferred setting of $\epsilon$ might be located in the range between 0.01 and 1.2}; \ding{205} \textbf{LDP is a better solution, compared with GDP, when deploying \DPabbr{} in the application-specific manner}; and \ding{206} \textbf{GDP and LDP protection on textual data can potentially destroy the performance of the whole application}. We summarize most significant observations, and discuss key takeaways from our findings in Section~\ref{sec:discussions}.

More specifically, we make three key contributions in this paper:

\begin{itemize}
    \item We address the challenges about preserving data privacy in the era of \concept{} (\concept{}), and identify key opportunities to tackle this challenge. We classify the applications of \conceptabbr{}, and outline the key mechanism \DP{} to enable privacy-preserving features of \concept{} in practice. 
    \item We design and characterize \DPabbr{}-enabled data processing and sharing on sensitive data, in the context of \conceptabbr{}. We investigate, implement and evaluate the state-of-the-art mechanisms from \DPabbr{}, over a state-of-the-art dataset for Human-Vehicle Interaction Research.
    \item Our characterizations derive FIVE most significant observations, selected from our 11 key findings. As our work is the first step to experimentally characterize the tradeoffs of privacy-preseving techniques for \conceptabbr{}, we believe our works are significantly beneficial for future research practices in privacy-preserving \conceptabbr{}.
\end{itemize}


\noindent
The rest of this paper is organized as follows. Section~\ref{sec:bkgd} introduces the background and motivation of our work. Section~\ref{sec:methodology} gives an overview of our study design regarding to the characterization of DP-enabled sensitive data processing and sharing.  Section~\ref{sec:processing} elaborates detailed characterization methodologies, and comprehensively analyzes the results. Section~\ref{sec:sharing} presents the experimental methodology for a real-world application with extensive results showcase and analysis. Section~\ref{sec:discussions} discusses several takeaways and future work.

\section{Background}
\label{sec:bkgd}
In this section, we introduce background and motivation of our characterizations. We first describe the background of \concept{} (Section~\ref{subsec:IoV}). Then we elaborate two major types of applications in \conceptabbr{} (Section~\ref{subsec:app-types}). Next, we introduce \DP{}, a general series of mechanisms to cover both types of \conceptabbr{} applications (Section~\ref{subsec:DP}). Finally, we identify our motivation and our goal in this work (Section~\ref{subsec:motivation}).

\subsection{The Concepts of \concept{}} 
\label{subsec:IoV}
\begin{figure*}
    \centering
    \includegraphics[width=0.8\linewidth]{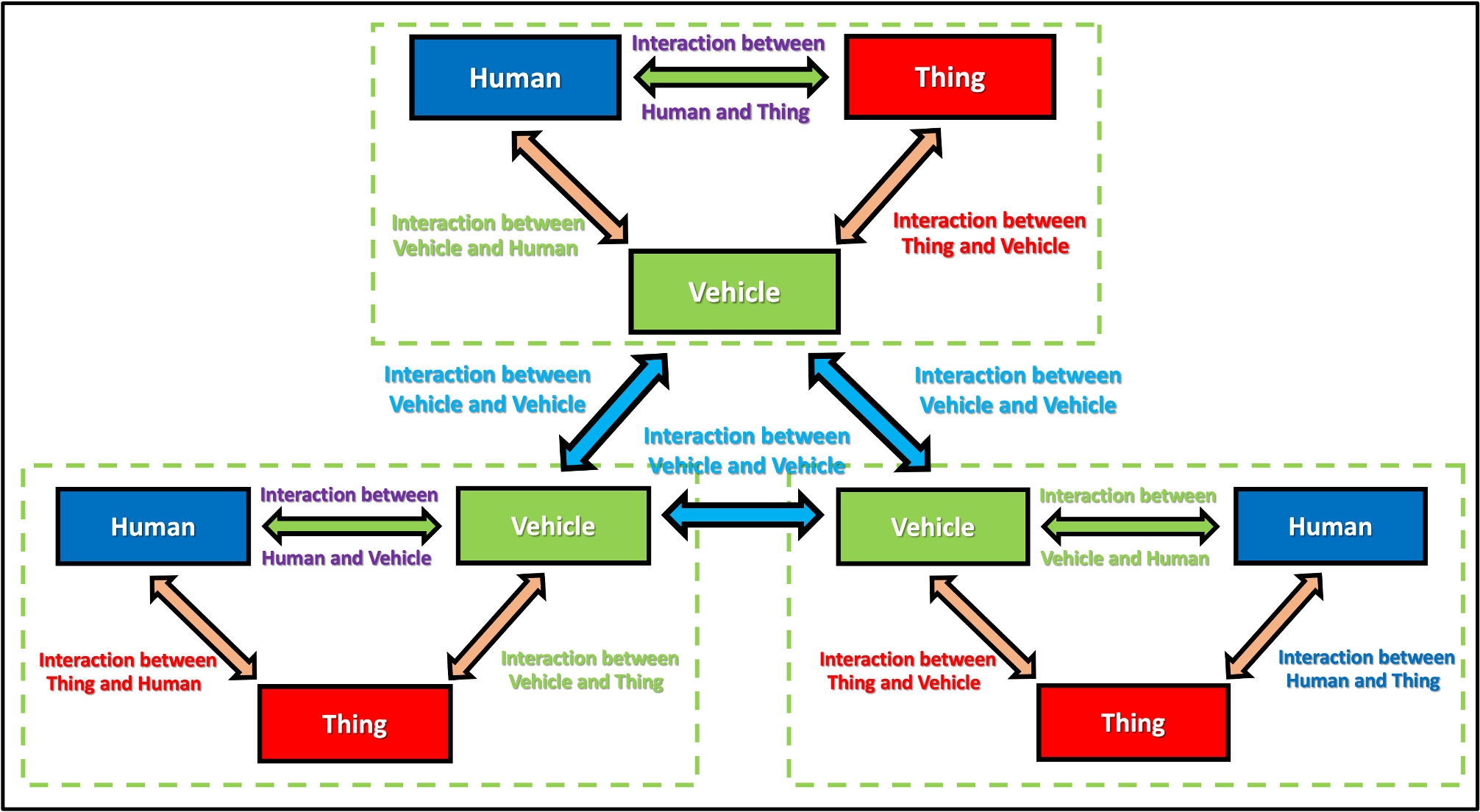}
    \caption{The architecture of \concept{}: (1) Human: people in vehicles and other people in the context of \conceptabbr{}, such as pedestrians; (2) Vehicle: all vehicles that consume or provide services/applications of \conceptabbr{}; (3) Thing: anything except human and vehicles, such as roads and remote servers;}
    \label{fig:conceptofIoV}
\end{figure*}

\concept{} (\conceptabbr{}) is a large-scale connected network for efficient cooperation's among humans, vehicles, things, and environments. Such cooperation consists of multiple interactions, which further provides sufficient and diverse services for a wide range of in-vehicle participants (e.g. city-wide or country-wide). As shown in Figure~\ref{fig:conceptofIoV}, the architecture of \conceptabbr{} can be abstracted as follow. \ding{202} All agents, including humans, vehicles, and things, connect with each other, within respective ranges of the particular agent types. These broad and extensive connections enable multiple use cases, such as traffic control \cite{traffic_control}, crash response \cite{crash_response}, convenience services \cite{18_IJARCSSE_introduction_of_IoV} and etc.; \ding{203} \conceptabbr{} also consists of many building blocks, with vehicles as the nodes, in practice. Within a \conceptabbr{}, each node of this network is a vehicle, and the vehicle is the primary unit of a complicated agent. All interactions from other agents (e.g. humans, things, etc.) have to go through the interfaces from vehicles, and then connect with others. It's notable to mention that, to manage a complicated architecture like \conceptabbr{}, centralized servers for groups of nodes are essential in practice. Therefore, though \conceptabbr{}'s architecture achieves tight associations among multiple agents (e.g., humans, vehicles, things and environments), this design also incurs potential leakage of sensitive personal statistics and substantially increases the risks of privacy disclosure.



\subsection{Classifying Application of \conceptabbr{}: Processing inside or outside the Vehicles?}
\label{subsec:app-types}

To better address the privacy concerns of \conceptabbr{}, we divide emerging applications of \conceptabbr{} into the following two types: Processing-In-Vehicles or Processing-Outside-Vehicles. We elaborate them in detail as follows. \ding{202} for Processing-In-Vehicles, there are some applications that can be fully processed within a vehicle. There are multiple examples in this scenario, such as the control of in-vehicle devices (e.g. conditions, lights, etc.), in-vehicle sensors for drivers \cite{In-Vehicle-Sensor} and in-vehicle monitors of driving statistics \cite{In-Vehicle_Monitor,automotive21/facial-expressions}; and \ding{203} for Processing-Outside-Vehicles, other applications may need assistance from centralized servers within \conceptabbr{}, due to either the application requirements or computational powers. There are many examples of applications, which require data aggregations in centralized servers. For instance, Transportation Scheduling demands all vehicles to update their coordinates and potential routines, to perform scheduling decisions \cite{Transportation_schedule}; also, there are many examples for applications, which demand more computational power in centralized servers (e.g., \cite{automotiveui21/iov, hci22/face2statistics,hci21/dbscan-driving-styles}). For instance, emerging trends of Deep Neural Networks demand significant computational power for model training, which are usually maintained in the centralized servers. Therefore, we consider our selected privacy-preserving techniques for \conceptabbr{} shall be applicable for both types of applications.


\subsection{Differential Privacy}
\label{subsec:DP}

\DP{} (\DPabbr{}) is a series of general privacy-preserving techniques, which are mathematically guaranteed for users' privacy, regardless of any auxiliary knowledge an adversary might have. The primary design concept of this effective mechanism is, to narrow the possible statistical effects from the increment of a single individual's data. Formally speaking, this is achieved by performing randomized functions subject to $\epsilon$-\DP{} on a database, to obtain the desired query outcome. The following definition gives out the mathematical description of a randomized function $K_f$.

\noindent
\begin{definition}
    (Randomized Function subject to $\epsilon$-\DP{}) We regard a randomized function $K_f$ as satisfying $\epsilon$-\DP{}, assume that $\forall$ dataset $D_1$ and $D_2$ with at most one element difference, and $\forall S \subseteq Range(K_f)$, the $K_f$ function obey the below in-equation:\\
    \centerline{$Pr[K_f(D_1) \in S] \leq exp(\epsilon) \times Pr[K_f(D_2) \in S]$}
\end{definition}

Note that, in practice, the realization of \DPabbr{} (Laplace Mechanism) usually depends on adding random noise, drawn from Laplace Distribution. The scale parameter is set as $\Delta f/\epsilon$, for $f(D)$ (i.e. where $D$ is the database, and $f$ is a query function): \ding{202} Three parameters $\epsilon$ should be manually selected at the beginning, according to the privacy protection degree the user anticipates. In other words, the smaller the $\epsilon$ is, the more noises sensitive data are added; and \ding{203} The other parameter $\Delta f$ ($L1-sensitivity$) relates to the query function $f$ to be used. After this noise addition, the query $f(D)\,+\,Laplace\,noise$ can achieve the same effects, as applying randomized function $K_f$ on $D$. In other words, the conceptual $\epsilon$-\DP{} function $K_f$ is instantiated by this approach. In this work, our focus lies on how to ensure our privacy protection meets the $\epsilon$-\DP{} standard. 

The reason why we select \DP{} as the core mechanism is that: \DP{} can be deployed for both Processing-In-Vehicle and Processing-Outside-Vehicles applications. We denote Processing-In-Vehicles applications as \textit{data processing of sensitive data}, and Processing-Outside-Vehicles as \textit{data sharing of sensitive data}. This features make \DP{} stands out from other privacy-preserving mechanisms, which are application-specific (e.g. Oblivious Inference \cite{ICML16/OI,arxiv18/FastOI}, Authentication \cite{TVT20/authentication-blockchain,IOTJ21/authentication-protocol,TITS20/authentication-grid} and etc.). Therefore, we believe \DPabbr{} is an ideal candidate for \conceptabbr{}, and it's worth the efforts to characterize its tradeoffs in detail.

\subsection{Motivations and Our Goal}
\label{subsec:motivation}

Though \DPabbr{} is a promising solution for privacy concerns of \conceptabbr{} in theory, it's still unclear how \DPabbr{} can introduce tradeoffs in the context of \conceptabbr{}. Obtaining experimental results, by applying \DPabbr{} in \conceptabbr{}, is the first step towards practical and effective solutions to privacy-preserving \conceptabbr{}. Our goal in this work is to investigate representative mechanisms of \DPabbr{} for both processing and sharing on sensitive data, and experimentally characterize the advantages and disadvantages in the context of \conceptabbr{}. Our study consists of two parts, focusing on data processing and data sharing on sensitive data respectively.

\section{Characterization Methodology}
\label{sec:methodology}

In this section, we describe our methodology to characterize differentially-private techniques for \conceptabbr{}. We first recap the rationales of choosing \DP{} as the core mechanism, based on the background of \conceptabbr{} (Section~\ref{subsec:rationales-DP}). Then we cover the design of our characterization, which is divided into two parts: data-processing and data-sharing (Section~\ref{subsec:study-design}). Next, we introduce the hardware, software and dataset used in our characterizations in detail (Section~\ref{subsec:HW-SF-Dataset}). Finally, we describe our efforts to emulate \conceptabbr{} in distributed machines (Section~\ref{subsec:emulation}).

\subsection{Rationales for \DP{}}
\label{subsec:rationales-DP}
The underlying reasons for choosing \DP{} as the core focus are that: (1) \DPabbr{} is one of the most effective and prevalent modern privacy-preserving techniques, which can be generalized for various applications for processing sensitive data; and (2) \DPabbr{} supports dynamic parameter tuning and several derivatives for context-specific purposes, which can also be specialized for specific applications for sharing sensitive data.

Since our characterization aims to obtain lessons in a relatively broad range, we believe the focus shall lie on a mechanism, which can be utilized in both application scenarios. Therefore, we believe \DPabbr{}-enabled data processing and sharing would become more important in \conceptabbr{}, and the characterization is an essential step to stimulate more studies for privacy-preserving applications in the era of \conceptabbr{}.


\subsection{Study Design}
\label{subsec:study-design}


To examine the tradeoffs of using \DP{} in the era of \concept, we divide our study into two separate parts for comprehensive characterizations. This is motivated by our classifications on \conceptabbr{} applications, which can be divided into data-processing and data-sharing on sensitive data. We describe our study design for separate parts in details as follows.

The first part of our characterizations focuses on \DPabbr-enabled sensitive data processing. We achieve so through three steps. First, we leverage existing database systems to add random noises to the query operators via \DPabbr{}. In this manner, querying driving physiological statistics (i.e. Heart Rate and Skin Conductivity) can be considered as abstracted processing in \conceptabbr{}. Second, we leverage four state-of-the-art \DPabbr{}-enabled queries (i.e., Fourier, Wavelet, DataCube and Hierarchical), to perform queries from real-world query workloads. To cover representative queries, we consider both All-Range Query and One-Way Marginal Query. Third, we evaluate the service quality of these queries, by using two distinctive metrics (i.e., Absolute Error and Relative Error compared with queries over the raw data). In this way, we can obtain key insights on how \DPabbr-enabled processing performs on sensitive data.

The second part of our characterizations emphasizes the \DPabbr-enabled sharing of sensitive data. We achieve so via three steps. First, we leverage a state-of-the-art application for in-vehicle drivers called Face2Multi-modal, a Deep-Neural-Network-driven predictor for physiological and sensitive statistics. Second, we identify key differences while training Face2Multi-modal using sensitive data, between centralized training and decentralized training. Third, we break down \DPabbr{} mechanisms into Global \DPabbr{} and Local \DPabbr{}, for centralized and decentralized training respectively. Fourth, we evaluate the service quality of both approaches, by using validation accuracy (i.e. how accurate the predictor is), compared with predictors over the raw data. In this way, we can retrieve key lessons on how \DPabbr{}-enabled sharing works on sensitive data.

\subsection{Hardware, Software and Dataset Supports}

\label{subsec:HW-SF-Dataset}
In this section, we cover details of the hardware, software and dataset supports for our characterizations. We first describe our hardware and software configurations. Then we elaborate the dataset for our studies in details.

\noindent
\ding{202} \textbf{Hardware Supports.} Our characterizations and emulations of \conceptabbr{} are performed on machines, whose detailed configurations are listed as follows. The CPU is Intel(R) Core(TM) i5-7200U and the GPU is NVIDIA GeForce MX150. We assume this configuration fits with a standard laptop/table-scale, which can be easily equipped for modern and emerging (autonomous) vehicles.

\noindent
\ding{203} \textbf{Software Supports.} The implementations of user-level applications (i.e. representative queries \& Face2Multi-modal) are implemented in Python. We use Python because it's very expressive and powerful to be integrated with modern database systems. In this manner, we can reduce as many variations of software configurations as possible, to ensure the consistency of our two studies.

\begin{figure*}
    \centering
    \includegraphics[width = \textwidth]{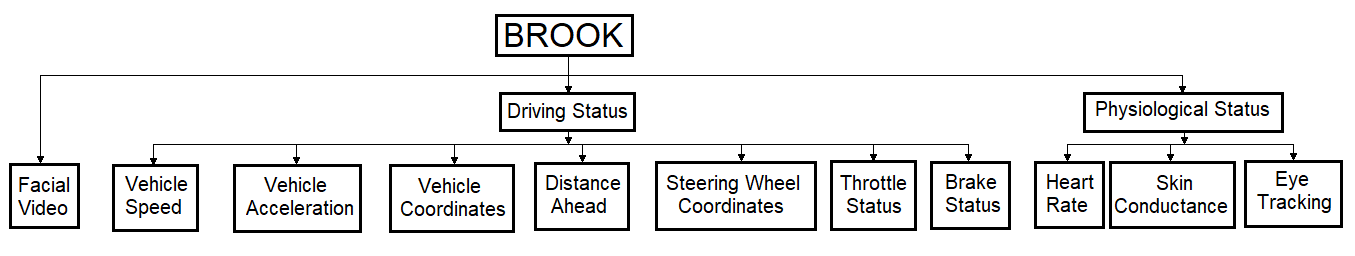}
    \caption{The composition of the BROOK dataset.}
    \label{fig:brook-structure}
\end{figure*}

\noindent
\ding{204} \textbf{Dataset Supports.} We use BROOK dataset for our characterizations \cite{peng2020building}. BROOK dataset has facial videos and multiple multi-modal/driving status data of 34 drivers. Statistics of every participant are collected in both manual mode and automated mode. Figure~\ref{fig:brook-structure} shows 11 dimensions of data for the current BROOK dataset, which includes: Facial Video, Vehicle Speed, Vehicle Acceleration, Vehicle Coordinate, Distance of Vehicle Ahead, Steering Wheel Coordinates, Throttle Status, Brake Status, Heart Rate, Skin Conductance, and Eye Tracking. 


\subsection{The Emulation of \concept{}}
\label{subsec:emulation}

We emulate \conceptabbr{} via the remote APIs of modern distributed database systems. This ensures our characterizations not to mistake any procedure, by overlooking the impacts of \conceptabbr{}. Since our focus lies on the effects on service quality, by enabling \DPabbr{} on both processing and sharing of sensitive data, we expect and validate our emulation to be sufficiently accurate, and there are no side effects on the presented results in the following sections.


\section{Characterizing \DPabbr{}-enabled Processing of Sensitive Data}
\label{sec:processing}
For \DPabbr{}-enabled processing, centralized servers perform queries to retrieve sensitive data from vehicle nodes, and then perform the computation within servers. In our characterizations, we select FOUR representative mechanisms for \DPabbr{}-enabled processing, and design representative workloads according to these mechanisms. Then we characterize the effects of these techniques and analyze the results comprehensively, in terms of service quality.

In this section, we characterize the impacts of DP-enabled data processing, by abstracting the interactions between vehicle nodes and centralized servers. We first brief the representative workloads and queries, which we select from the state-of-the-art approaches (Section~\ref{subsec:query-types}). Next, we elaborate the setup for examining the impacts of representative queries (Section~\ref{subsec:query-setup}). Then, we describe how we construct various query structures and tune different $\epsilon$  values for broad coverage of our characterizations (Section~\ref{subsec:query-config}). Finally, we describe our results by characterizing the impacts of DP-protected data through the above two approaches, in terms of service quality (Section~\ref{subsec:query-results}). 

\subsection{\DPabbr{} Mechanisms for Processing}
\label{subsec:query-types}

In this section, we cover four state-of-the-art mechanisms for enabling \DPabbr{} on queries,  under $\epsilon$-\DP{}. We denote these mechanisms using words reflecting their key ideas, which are Fourier, Wavelet, DataCube and Hierarchical. We describe these mechanisms in details one-by-one.

\noindent
\ding{202} \textbf{Fourier.} The design purpose of Fourier is specifically for workloads having all K-way marginals, for any given $K$ \cite{07_SIGMOD_fourier}. This mechanism transforms the cell counts, via the Fourier transformation, and obtains the values of the marginals based on the Fourier parameters. If the workload is not full rank, there are eliminations of the unnecessary queries from the Fourier basis to reduce the sensitivity. 

\noindent
\ding{203} \textbf{Wavelet.} The design goal of Wavelet is to support multi-dimensional range workloads. This mechanism applies the Haar wavelet transformation to each dimension \cite{10_IEEE_Wavelet}. Note that Wavelet is expected to be applied in a hybrid manner if using $\epsilon$-differential privacy. Such a hybrid algorithm depends on the number of distinct values in the selected dimension: when the number is small, the identity strategy is applied to improve Wavelet's performance. 

\noindent
\ding{204} \textbf{DataCube} The target applicability of DabaCube is to adaptably support marginal workloads \cite{11_SIGMOD_datacube}. To provide a deterministic comparison point, the BMAX algorithm is applied in this case. This algorithm chooses a subset of input marginals, to minimize the maximum errors of the answer to the input workloads. DataCube is naturally adaptable with ($\epsilon$, 0)-differential privacy. 

\noindent
\ding{205} \textbf{Hierarchical} The aim of Hierarchical~\cite{hierarchical} is to answer workloads of range queries. The key idea is to use a binary tree structure of queries: (1) the first query is the sum of all cells; and (2) the rest of the queries recursively divide the first query into subsets. As the representative comparison point, we apply binary hierarchical strategies (although higher orders are possible). The strategy in \cite{hierarchical} supports one-dimensional range workloads, but is capable to be adapted to multiple dimensions, similarly to Wavelet \cite{10_IEEE_Wavelet}.

\subsection{Dataset Setup and Query Types}
\label{subsec:query-setup}


In this section, we aim to cover the concept definitions of All-Range Query and One-Way Marginal Query. Due to the close relationship between these two queries and the expression of our dataset, we will discuss the dataset set up at the beginning and then concentrate on defining these two kinds of counting queries.

\noindent
\ding{202} \textbf{Dataset Setup}
To characterize the query errors from the above types of \DPabbr{}-enabled query mechanisms, we rectify our dataset, to import them into our emulation platform. More specifically, we pair arbitrary types of data streams, in an evenly distributed manner, and iterate all possible cases with average results reported. We use the combination of Heart Rates and Skin Conductivity as a motivating example. Assume the table is of size $m \times n$, and the constraint condition for cell $(i, j)$ could be written as $\rho_{ij}$. Then, a $length-(m \times n)$ column vector $\mathbf{x} = (x_{00}, x_{01}, ... , x_{(m - 1)(n - 1)})^T$ where $x_{ij}$ meets the condition $\rho_{ij}$ could be extracted from the table. In this case, it is equal to $(12, 17, 0, 0, 2254, ..., 11)^T$, a column vector with length of 32. With the existence of this data vector $\mathbf{x}$, it is convenient to answer a linear query. 

\noindent
\ding{203} \textbf{All-Range Query} All-Range Query, defined in this article, is closely associated with the data vector. The dot product of a single $length-n$ row vector, with (1) all positions being $1$ and (2) a $length-n$ column data vector, is the expected result of the All-Range Query result. For a database whose content is certain and unchanged, performing All-Range Query would always generate the same results. 

\noindent
\ding{204} \textbf{One-Way Marginal Query} One-Way Marginal query is a concept derived from the K-Way Marginal query. We first provide a general explanation regarding the K-Way Marginal query. A K-Way Marginal Query on a $n$-dimension database expects a returned vector, whose each position corresponds to the counting result of a query related to partial attributes (k attributes) only. Also, the magnitude of all possible settings considering $K$ attributes is the length of this vector. Similarly, the One-Way Marginal query is a special case of k-Way Marginal query when $K=1$.

\subsection{Metrics and Configurations of Workloads}
\label{subsec:query-config}
\noindent
\ding{202} \textbf{Absolute Error.} The absolute error refers to the absolute difference between the measured value and real value. In our experiment, we regard the query outcomes generated by four \DPabbr{}-related query approaches as measured values. And corresponding queries without \DPabbr{} processing directly performed on the database will give rise to real values. For instance, given that an All-Range Query utilizing Wavelet method obtains result $r'$ when the dimension setting is $4\times8$, and, in the same condition, direct All-Range Query on the database will get $r$, the absolute error, in this case, will be $|r'-r|$. Note that, when $r$ and $r'$ are all vectors, the absolute error becomes $\|r'-r\|_1$.

\noindent
\ding{203} \textbf{Relative Error.} The relative error is defined as the ratio of the absolute error to the real value. The formal mathematical expression for it is $Relative\,Error = Absolute\,Error/Real\,Value$. Considering the afore-mentioned example, the relative error in that situation should be $|(r'-r)/r|$.

\noindent
\ding{204} \textbf{Varied Configurations.} In experiments for absolute error, the parameter $\epsilon$ is previously fixed to $0.5$, and the dimension setting is the only changeable variable, which rigorously follows the control variable method. Concerning experiments related to the relative error, they are all conducted on a $4\times8$ database picked from BROOK with some study-worthy $\epsilon$ values (i.e. 0.1, 0.2, 0.5, 1 and 2.5) for the adjustment of four \DPabbr{} query methods.

\noindent
\ding{205} \textbf{Workloads for All-Range Query} According to the definition of All-Range Query, the desired counting result shall cover all possible constraint conditions appearing in data vector, which, specifically speaking, for any linear query in the workloads of All-Range Query, it is supposed to be a vector with all positions be set as one, regardless of dataset dimensions. \\
\noindent
\ding{206} \textbf{Workloads for One-Way Marginal Query} As for one-way marginal query, we rigorously examine the query results on various dimension settings (i.e. $32$, $4\times8$, $4\times4\times2$, $4\times2\times2\times2$ and $2^5$). Various One-way Marginal Queries regarding different data types such as HR and SC could be realized through linear queries. By integrating all the linear queries with different structural designs into one matrix, we will obtain the one-way marginal query workload.





\subsection{Characterizations Results}
\label{subsec:query-results}
\begin{figure}[htp]

\centering

    \begin{minipage}{0.98\columnwidth}
        \centering
        \includegraphics[width=0.98\columnwidth]{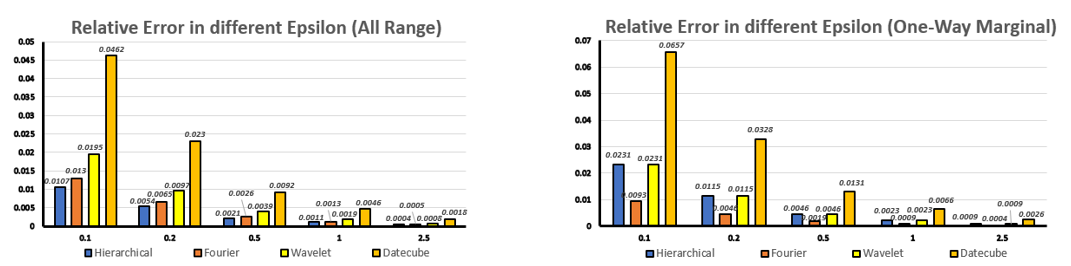}
        \caption{Relative Error of Four Query Methods}
        \label{fig:relative}
    \end{minipage}

    \begin{minipage}{0.98\columnwidth}
        \centering
        \includegraphics[width=0.98\columnwidth]{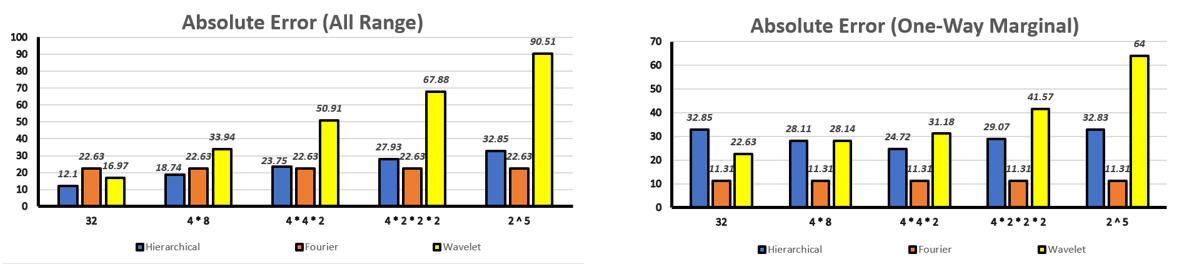}
        \caption{Absolute Error of Four Query Methods}
        \label{fig:absolute}
    \end{minipage}

\label{fig:}

\end{figure}

In this section, we demonstrate the query results of the four representative mechanisms under our settings, in terms of Relative Error and Absolute Error. Figure~\ref{fig:relative} depicts the Relative Errors for the covered four mechanisms, for both All-Range Query and One-Way Marginal Query. We make three key observations: First, for both All-Range Query and One-Way Marginal Query, DataCube incurs more errors than any other mechanisms in all scenarios, which suggests it may not be an ideal candidate in practice. Second, Hierarchical is the best query method for all the cases in All-Range Query, but Fourier performs the best for every $\epsilon$ parameter setting in One-Way Marginal Query. Therefore, we conclude that \textbf{there are no one-size-fits-all mechanisms of \DPabbr{} for all applications of \conceptabbr{}, in terms of Relative Error (i.e. Key Observation 1)}. Third, all mechanisms gradually reduce their Relative Errors, when the $\epsilon$ increases; moreover, all mechanisms achieve the best query results at the maximum $\epsilon$ setting. 

Figure~\ref{fig:absolute} shows the results for Absolute Error of the covered four mechanisms. Note that DataCube is incapable to evaluate Absolute Errors since this approach is highly dataset-dependent. We make four key observations. First, the query error of both Hierarchical and Wavelet has a positive correlation with the growing granularity of query structures in both query types. Second, the query errors of Fourier are more robust with the growing granularity of query structures. Particularly, Fourier achieves the best results in complex granularity of query structures for All-Range Query, and it also achieves the best in all cases for One-Way Marginal Query. Third, for the query structure $32$, $4 \times 8$ in All-Range Query, Hierarchical obtains the best query results rather than Fourier. Fourth, compared with Hierarchical, Wavelet can achieve less Absolute Error in some simpler combinations for One-Way Marginal Query. Therefore, we conclude that \textbf{there are no one-size-fits-all mechanisms of \DPabbr{} for all applications of \conceptabbr{}, in terms of Absolute Error as well (i.e. Key Observation 2)}.

\section{Characterizing \DPabbr{}-enabled Sharing of Sensitive Data}
\label{sec:sharing}

In this section, we characterize the impact of DP-enabled data sharing, by focusing on a novel in-vehicle application, Face2Multi-Modal \cite{huang2020face2multi}. We first brief the background of Face2Multi-Modal, and discuss the potential privacy leakages in either centralized or decentralized manners (Section~\ref{subsec:f2m-background}). Next, we elaborate on two approaches to resolve privacy leakages in centralized and decentralized manners, using \DPabbr{} (Section~\ref{subsec:GDP-LDP}). Then, we vary the configurations of GDP and LDP, to enable a broad coverage of our characterizations (Section~\ref{subsec:config-f2m}). Finally, we describe our results by characterizing the impacts of DP-protected data through the above two approaches, in terms of service quality (Section~\ref{subsec:DP-f2m}).

\subsection{Face2Multi-modal: A Motivating Example for Privacy Concerns of In-Vehicle Applications}
\label{subsec:f2m-background}

Face2Multi-Modal provides an in-vehicle multi-modal data stream predictor, by taking facial expressions as the only input \cite{huang2020face2multi}. Face2Multi-modal is motivated by the fact that: obtaining drivers' physiological data, by directly equipping bio-sensors, is expensive, impractical and user-unfriendly. The key idea of Face2Multi-modal is to utilize the Convolutional Neural Networks as the backbone for predicting driver statistics. 

Since, for training, both input (i.e. facial expressions) and output (i.e. drivers' statistics like Heart Rate, Skin Conductance, etc.) are highly sensitive information, the deployments of Face2Multi-modal can incur significant concerns in terms of privacy. Moreover, the service quality of Face2Multi-modal is highly related to the privacy-preserving techniques, and the reasons are two-folded. First, the training of Face2Mutli-modal is computationally expensive, and it's ideal to perform such training on centralized servers. and second, the service quality of Face2Multi-modal relies on the model training on a high volume of raw data, and applying privacy-preserving techniques can increase the difficulty of getting accurate models.

In this characterization, we consider two types of model training for Face2Multi-modal and enable \DPabbr{} for them using corresponding mechanisms. We first consider that Face2Multi-modal is trained in a centralized manner. More specifically, a training of Face2Multi-modal is performed based on aggregated data received from different nodes and processed in the centralized servers. Then we consider that Face2Multi-modal is trained in a decentralized manner, where the training of Face2Multi-modal is performed solely in a vehicle node, where no data are aggregated into the centralized servers.

\subsection{\DPabbr{}-enabled Training of Face2Multi-modal: Two Approaches }
\label{subsec:GDP-LDP}

To enable \DPabbr{}-protected training of Face2Multi-modal, we leverage Global \DP{} (GDP) for centralized training, and Local \DP{} (LDP) for decentralized training on sensitive data respectively. Hereby, we provide sufficient details regarding these two approaches, via elaborating key definitions and differences as follows.

\noindent
\ding{202} \textbf{Global \DP{} (GDP).} GDP manages data for a centralized scenario. In GDP, the application has a central server as the aggregator, which accepts data from vehicle nodes. In this way, the aggregator (i.e. centralized servers in \conceptabbr{}) can have the access to all raw data, and therefore all data are exposed to the aggregator. The idea of GDP is to allow the aggregator to collect all sensitive data from vehicle nodes, and apply a \DPabbr{} mechanism to add random noise on the whole aggregated dataset. In this way, GDP can protect all sensitive data by adding the noise globally, thus reducing the possibility of privacy leakage when the attacker obtains the same privilege as the aggregator. 

\noindent
\ding{203} \textbf{Local \DP{} (LDP).} LDP protects data in a decentralized scenario. In contrast to GDP, LDP protects data on the client side. As edge devices could collect extensive data about users, end-users might confront the privacy leakage problems when an attacker cracks their devices and controls the access to their sensitive data. LDP protects users' data by adding random noise locally to provide safety guarantees. The processed sensitive data would then be used for data sharing either locally or among the Vehicle Network.

We use GDP and LDP for centralized and decentralized training respectively. For centralized training, GDP protection works by 1) collecting all drivers' statistics in centralized servers; 2) adding noises using \DPabbr{} on the gathered data set; and 3) performing training for Face2Multi-modal. As for decentralized training, LDP protection works by 1) adding noises to individual drivers' statistics in-vehicle nodes; and 2) performing training for Face2Multi-modal.

\subsection{Configurations of Our Characterizations}
\label{subsec:config-f2m}

We vary the configurations during the training of Face2Multi-modal. Specifically, we consider three aspects of our experiments. First, we vary the range of $\epsilon$  (i.e. $\epsilon$ reflects the extent of noises on raw data), when applying GDP and LDP. Then, we vary the objective types for \DPabbr{}-enabled protection, including both images (i.e. input for Face2Multi-modal) and textural information (i.e. output of Face2Multi-modal).

\noindent
\ding{202} \textbf{GDP Parameters.} Our emulation aggregates all data together, namely $F=F_{1}+...+F_{N}$ , $S=S_{1}+...+S_{N}$, where $F$ and $S$ are fed into the training of Face2Multi-Modal. During this procedure, we vary the value of epsilon (i.e. $\epsilon$). Therefore, we can get different $F$ and/or $S$, as the source data, to perform the training of Face2Multi-modal.

\noindent
\ding{203} \textbf{LDP Parameters.} Our emulation considers the whole procedure that is conduced  within each vehicle node under LDP protection. Therefore, each vehicle node has its own $F$ and $S$, adds their own noises and trains their own models. Similarly, we also vary the value of epsilon as we did in GDP parameters.

\noindent
\ding{204} \textbf{Protected Data Types.} Our characterizations focus on the data types, suggested by Face2Multi-modal. We consider raw data for training, including both input and output for training, can be protected by \DPabbr{}. We enable the protection on reasonable combinations of images and textual information.

\subsection{Characterization Results}
\label{subsec:DP-f2m}
In this section, we present the characterization results of GDP and LDP protection approaches on Face2Multi-Modal. First, we visualize the $\epsilon$-\DPabbr{} protection of image data in different settings of $\epsilon$. Next, we showcase the validation accuracy of Face2Multi-modal, in terms of textual information, for GDP and LDP approaches, with protecting drivers' facial images only. Finally, we present the validation accuracy of Face2Multi-Modal when both textual and images are protected under \DPabbr{} simultaneously.

\begin{figure*}[!h]
\minipage{0.48\textwidth}
  \centering
    \includegraphics[width=0.8\linewidth]{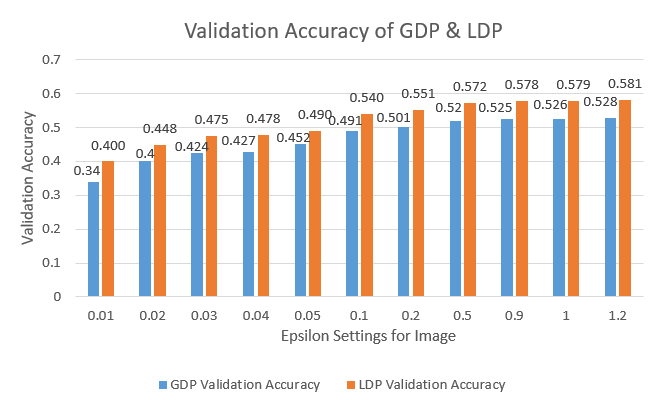}
    \caption{Validation Accuracy between GDP and LDP in different Epsilon settings for facial images.}
    \label{fig:f2m_result}
\endminipage\hfill
\minipage{0.48\textwidth}%
    \includegraphics[width=0.8\linewidth]{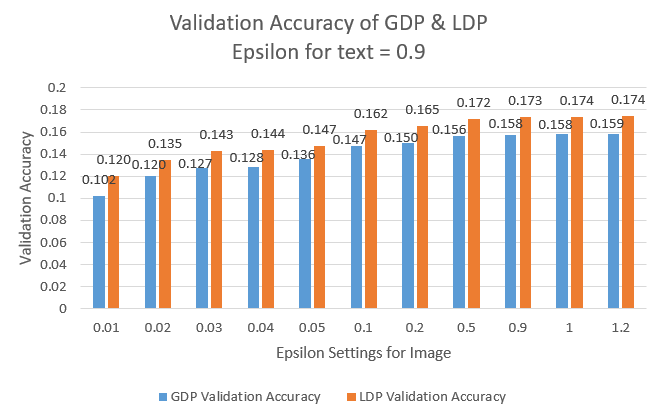}
    \caption{Validation Accuracy between GDP and LDP in different Epsilon settings for textual information.}
    \label{fig:f2m_result_txt}
\endminipage\hfill
\end{figure*}

Figure~\ref{fig:f2m_result} presents the validation accuracy for GDP and LDP in various $\epsilon$ settings. We make two key observations. First, LDP outperforms GDP in the Face2Multi-Modal system by an average of 1.7X improvement of estimation accuracy. This is because Deep Neural Network Models on each vehicle are trained locally and customized for specific drivers. \textbf{This suggests that LDP is a better solution, compared with GDP, when deploying \DPabbr{} in the application-specific manner (i.e. Key Observation 3). } Second, the validation accuracy of both approaches keeps growing from the lowest $\epsilon$ value to the highest value, since high $\epsilon$ value adds relatively low noises on the raw data. Therefore, the validation accuracy is close to the validation accuracy, when training on the raw data. 

Figure~\ref{fig:f2m_result_txt} showcases the Validation Accuracy of GDP and LDP after introducing $\epsilon$-\DP\;to protect textual information (i.e. $\epsilon$=0.9) on the basis of various $\epsilon$ settings for image protection. By comparing the results in Figure~\ref{fig:f2m_result_txt} and Figure~\ref{fig:f2m_result}, we can conclude that both \textbf{GDP and LDP protection on textual data can potentially destroy the performance of the whole application (i.e. Key Observation 4)}, by reducing the validation accuracy to one thirds of the previous one.

\begin{figure}
    \centering
    \includegraphics[width=0.9\linewidth]{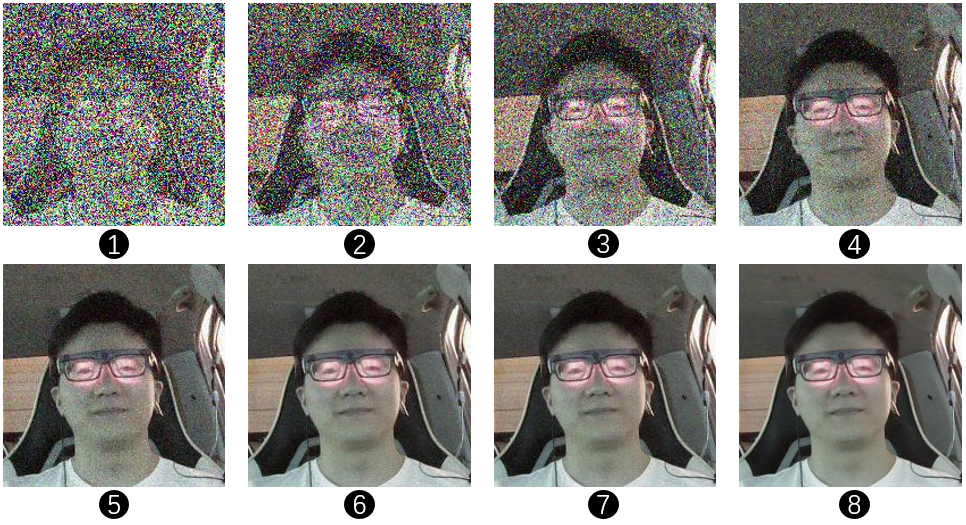}
    \caption{Different $\epsilon$ settings and corresponding visual effects. For each image, the $\epsilon$ values are \ding{202} $\epsilon$ =0.005; \ding{203} $\epsilon$ =0.01; \ding{204} $\epsilon$ =0.02; \ding{205} $\epsilon$ =0.05; \ding{206} $\epsilon$ =0.1; \ding{207} $\epsilon$ =1.}
    \label{fig:epsilon-setting}
\end{figure}

Figure \ref{fig:epsilon-setting} shows the visual effects of the same image under different $\epsilon$ settings. From visual inspection, the extent of noises, added into the image, is in inverse proportion to the values of $\epsilon$. For \ding{202}, the facial image is "over-protected", which indicates that the random noises might destroy the inherent biological facial features used for Deep-Neural-Network model inference thus degrading the overall performance. While for \ding{209}, it has a relatively high $\epsilon$ value, which protects the image with minimum effort. Therefore, \textbf{to strike a balance between data integrity and privacy preserving, a preferred setting of $\epsilon$ might be located in the range between 0.01 and 1.2  (i.e. Key Observation 5)}.  

\section{A Summary of The Most Significant Observations \& Our Takeaways}
\label{sec:discussions}



In this section, we summarize our FIVE significant observations, derived from our 11 key findings through the rigorous characterizations, and discuss our takeaways from them. We first describe our most significant observations in detail (Section~\ref{subsec:observations}). Then we elaborate our takeaways from these observations and suggest both challenges and opportunities for privacy-preserving \conceptabbr{} (Section~\ref{subsec:takeaways}).

\subsection{FIVE Most Significant Observations}
\label{subsec:observations}
We summarize our key observations into the following five aspects, derived from two separate characterizations (Note that \DPabbr{} refers to \DP{}, \conceptabbr{} refers to \concept{}, GDP refers to Global \DP{}, LDP refers to Local \DP{}): 

\noindent
\ding{202} \textbf{There are no one-size-fits-all mechanisms of \DPabbr{} for all applications of \conceptabbr{}, in terms of Relative Error}.

\noindent
\ding{203} \textbf{There are no one-size-fits-all mechanisms of \DPabbr{} for all applications of \conceptabbr{}, in terms of Absolute Error};.

\noindent
\ding{204} \textbf{To strike a balance between data integrity and privacy preserving, a preferred setting of $\epsilon$ might be located in the range between 0.01 and 1.2}.

\noindent
\ding{205} \textbf{LDP is a better solution, compared with GDP, when deploying \DPabbr{} in the application-specific manner}.

\noindent
\ding{206} \textbf{GDP and LDP protection on textual data can potentially destroy the performance of the application}.

We conjecture that these findings have covered all major design issues, when applying \DPabbr{} in the context of \conceptabbr{}.

\subsection{Takeaways of Our Characterizations}
\label{subsec:takeaways}

We discuss three major takeaways of our characterizations, based on the most significant observations. First, since there are no one-size-fits-all solutions for privacy-preserving \conceptabbr{}, by using DP. We believe there are a large number of opportunities to explore privacy-preserving techniques for \conceptabbr{} in practice. Second, for data processing on sensitive data, we already narrow the large design scope into a restricted region, by examining the value of $\epsilon$ in diverse settings. We believe this is very useful for follow-up studies, to avoid redundant workloads of attempted settings. Third, for data sharing on sensitive data, we already demonstrate that application-specific \DPabbr{} protections can potentially destroy the service quality. Therefore, we believe there are a huge amount of possibilities to co-design the application and privacy-preserving mechanisms (e.g., DP-based protection for unbalanced batch queries in IoV \cite{UCC-TR/HUT}), to ensure high service quality with privacy protection at the same time.

\section{Conclusions}
We present the most comprehensive study to characterize modern privacy-preserving techniques for \conceptabbr{}. Our study provides 11 key findings for applying \DPabbr techniques in the context of \conceptabbr and we highlight FIVE most significant observations from our detailed characterizations. We focus on \DP{} (\DPabbr{}), a representative set of mathematically-guaranteed mechanisms for both privacy-preserving processing and sharing on sensitive data. The purpose of our study is to demystify the tradeoffs of leveraging \DPabbr{} techniques, with service quality as the evaluation metrics. We first characterize four representative privacy-preserving processing, enabled from advanced \DPabbr{}. Then we perform a detailed study of an emerging in-vehicle, Deep-Neural-Network-driven application, and study the strength and weakness of \DPabbr{} for diverse types of data streams. We conclude that there is a large volume of challenges and opportunities for future studies, by enabling privacy-preserving \conceptabbr{} with low overheads for service quality.


\bibliographystyle{ACM-Reference-Format}
\bibliography{sample-base}

\appendix

\end{document}